# New Iron-based arsenide oxides (Fe$_2$As$_2$)(Sr$_4$M$_2$O$_6$)(M = Sc, Cr)

Hiraku Ogino[1,2], Yukari Katsura[1], Shigeru Horii[1,2], Kohji Kishio[1,2] and Jun-ichi Shimoyama[1,2]

[1]Department of Applied Chemistry, The University of Tokyo, 7-3-1 Hongo, Bunkyo-ku, Tokyo 113-8656, Japan
[2]JST-TRIP, Sanban-cho, Chiyoda-ku, Tokyo 102-0075, Japan
e-mail address: tuogino@mail.ecc.u-tokyo.ac.jp

## Abstract

We have discovered new arsenide oxides (Fe$_2$As$_2$)(Sr$_4$M$_2$O$_6$) (M = Sc, Cr: M-22426). These materials are isostructural with (Fe$_2$P$_2$)(Sr$_4$Sc$_2$O$_6$), which was found in our previous study. The new compounds are tetragonal with a space group of *P4/nmm* and consist of the anti-fluorite type FeAs layer and perovskite-type blocking layer. These compounds have long interlayer Fe-Fe distances corresponding to the *c*-axis length, the 15.8 Å in Sc-22426 is the longest in the iron-based pnictide oxide systems. Chemical flexibility of the perovskite block in this system was probed by chromium containing (Fe$_2$As$_2$)(Sr$_4$Cr$_2$O$_6$). Different trends were found in bond angle and bond length of the new pnictide oxides compared to the reported systems, such as *RE*Fe*Pn*O. Absence of superconductivity in these compounds is considered to be due to insufficient carrier concentration as in the case of undoped *RE*FeAsO.

**PACS:** 61.05.cp X-ray diffraction, 61.66.Fn Inorganic compounds, 74.70.-b Superconducting materials

**Keyword:** pnictide oxide, superconductor, perovskite structure

## Introduction

High-$T_c$ superconductivity in iron based pnictide oxide systems [1] has proposed new guidelines for development of superconducting materials. Several types of iron-based superconductors are discovered such as LiFeAs [2], *AE*Fe$_2$As$_2$(abbreviated as 122, *AE* = Alkali earth metals) [3], *RE*FeAsO(abbreviated as 1111, *RE* = Rare earth elements) [4], *AE*FeAsF [5] and related phosphide oxide and chalcogenide materials. Meanwhile, still there are continuous demands for new materials containing iron tetragonal lattice particularly for achieving higher $T_c$.

Several discussions have been already made for determining factors of $T_c$ in iron-based pnictide oxides. High symmetry in Fe$Pn_4$ tetrahedra is pointed out to be an important factor for high $T_c$. Actually, the highest $T_c$ for 1111 and 122 phases were achieved by compounds with angles between As-Fe-As of ~109.5°, which is the desirable value for perfect symmetry. In addition, from chemical point of view, the high $T_c$ exceeding 30 K under ambient pressure was achieved only in compounds having FeAs layer. Combination of iron and arsenic might be one of the key factors to achieve high $T_c$. In our recent study [6], we have discovered a new superconducting pnictide oxide $(Fe_2P_2)(Sr_4Sc_2O_6)$ with $T_c \sim 17$ K. This material has an alternate stacking of perovskite ($K_2NiF_4$-type) oxide layer and anti-fluorite pnictide layer. Flexibility of the perovskite-based structure in sulfide oxide and pnictide oxide systems is already investigated and several structure types and variety of constituent transition metals in perovskite layer including FeAs-based material have been reported [7-14]. These motivated us to search new iron pnictide oxides with perovskite-type oxide layers. In this study, we have extended the previous discovery of $(Fe_2P_2)(Sr_4Sc_2O_6)$ to the arsenide oxide system. New iron arsenide oxides $(Fe_2As_2)(Sr_4M_2O_6)$ ($M$ = Sc, Cr) (abbreviated as $M$-22426) have been successfully synthesized and their structural and physical properties were characterized.

**Experimental**

All samples were synthesized by solid-state reaction starting from FeAs(3N), SrO(2N), Sr(2N), Cr(3N), $Sc_2O_3$(4N) and $Cr_2O_3$(3N). Nominal compositions were fixed according to this general formula : $(Fe_2As_2)(Sr_4M_2O_6)$ ($M$ = Sc, Cr). Since the starting reagents, Sr and SrO, are sensitive to moisture in air, manipulations were carried out under an inert gas atmosphere. Powder mixture of reagents was pelletized and sealed in evacuated quartz ampoules. Heat-treatments were performed in the temperature range from 1000 to 1200°C for 24 to 50 hours.

Phase identification was carried out by X-ray diffraction (XRD) with RIGAKU Ultima-IV diffractometer and intensity data were collected in the $2\theta$ range of 5° - 80° at a step of 0.02° using Cu-$K_\alpha$ radiation. Silicon powder was used for the internal standard. Structural refinement was performed using the analysis program RIETAN-2000[15]. High-resolution images were taken by JEOL JEM-2010F field emission TEM. Magnetic susceptibility measurement was performed by a SQUID magnetometer (Quantum Design MPMS-XL5s). Electric resistivity was measured by AC four-point-probe method using Quantum Design PPMS.

**Results&Discussions**

Bulk samples of Sc-22426 with single phase were obtained by sintering at 1200°C. On the other hand, phase purity of isostructural Cr-22426 samples was relatively poor even after sintering at 1200°C for a long time, though the target compound was formed as a main phase. Small amounts of FeAs, SrO and other unknown impurities were included in the samples. The powder XRD patterns of Sc-22426 and Cr-22426 reacted at 1200°C for 40 h are shown in Fig. 1. Lattice constants were determined to be $a$ = 4.050 Å and $c$ = 15.809 Å for Sc-22426 and $a$ = 3.918Å and $c$ = 15.683 Å for Cr-22426. Space group of the both compounds is *P4/nmm*. Similar to a Sc-22426 compound with FeP layer [6], the interlayer Fe-Fe distance of Sc-22426 with FeAs layer is very long ~15.8 Å, which is the longest value ever reported in the iron pnictide oxide systems.

Rietveld refinement was carried out for powder XRD patterns of Sc-22426 and Cr-22426. In the case of Sc-22426, there are no traces of impurity or other phases, such as $(Fe_2As_2)(Sr_3Sc_2O_5)$ (abbreviated as Sc-22325), which was found by Zhu *et al*.[14]. The difference of Sc-22426 and Sc-22325 is the stacking pattern of the perovskite structure. In the refinement for Cr-22426, diffraction peaks derived from impurity phases were excluded. The results of the refinement are summarized in Table 1, while *R* factors of Cr-22426 was rather poor because of coexisting several impurities. It was found that the *a*-axis length of Sc-22426 is slightly shorter and $\alpha$ angle is slightly smaller than those of Sc-22325. Meanwhile, Cr-22426 has a short *a*-axis and small $\alpha$ angle compared to Sc based iron pnictide oxides. XRD measurements were performed down to 15 K to investigate structural changes, however, no structural change was observed for either compounds.

Figure 2 shows a bright-field TEM image and an electron diffraction pattern taken from [1 -1 0] direction of a Sc-22426 crystal. Both TEM image and electron diffraction patterns indicated tetragonal cell with $c/a \sim 3.9$, which coincide well with a corresponding value obtained from XRD patterns. It should be noted that any stacking faults were not found in the observed crystals, whereas the perovskite block has large variety in layer stacking pattern, such as $Sr_3Sc_2O_5$. Furthermore, absence of satellite spots probes commensurate stacking between $Fe_2As_2$ and $Sr_4Sc_2O_6$ layers.

Temperature dependences of magnetization of Sc-22426 and Cr-22426 measured under 10 kOe are shown in Fig. 3. Curie-Weis like behavior without anomaly was observed for the Sc-22426, while Cr-22426 showed a broad magnetization peak around 80 K. This behavior is similar to that of an sulfide oxide analogue $(Cu_2S_2)(Sr_4Cr_2O_6)$[9]. In the case of $(Cu_2S_2)(Sr_4Cr_2O_6)$, this behavior is explained by antiferromagnetic ordering of $Cr^{3+}$ in the perovskite-type layer derived from spin 3/2 at Cr site. Both compounds did not exhibit any signs of superconductivity down to 1.7 K possibly due to insufficient carrier concentration as in the case of the undoped REFeAsO compounds.

Figure 4 shows temperature dependence of resistivity for Sc-22426 and Cr-22426. As in the case of Sc-22325 [14], any anomaly was not found in the resistivity curve. On the contrary, Cr-22426 showed metallic behavior above 130 K and anomalous drop below 60 K. Although origins of the anomaly are not clear at the present stage, it is interesting that the temperature of the anomaly in resistivity is slightly different from that in magnetization.

In order to clarify structural features of the perovskite-based pnictide oxides, $\alpha$ angles of Fe$Pn$ layer and distances between Fe and $Pn$ of reported phases are arranged by their $a$-axis length as shown in Fig. 5. Both $\alpha$ angles and Fe-$Pn$ distances tend to increase with increasing $a$-axis length to maintain the crystal structures. On the other hand, there are differences in trends derived from structural groups. For example, $\alpha$ angle of Sc-22426 is similar to that of Ce-1111 while $a$-axis of the compound (4.050 Å) is much longer than that of Ce-1111 (4.000 Å). This result is due to longer Fe-As distance in Sc-22426 than that in Ce-1111. As clearly seen in this figure, this is the general tendency between perovskite based arsenide oxides and $RE$-1111 phases. This situation resembles a relationship between LaFePO and $(Fe_2P_2)(Sr_4Sc_2O_6)$. $\alpha$ angle of $(Fe_2P_2)(Sr_4Sc_2O_6)$ is closer to angle of ideal Fe$Pn_4$ tetrahedron (109.47°) than that of LaFePO owing to its longer Fe-$Pn$ distance. These facts suggest that the perovskite-based iron pnictide oxides have different nature of local iron pnictide structure. Suggested by Lee et al. [20], the compounds with the less distorted FeAs$_4$ tetrahedron tends to exhibit higher $T_c$. The different trend in angles and bond length of pnictide layer in perovskite-based pnictide oxides will bring new ways to control local structures.

The discovery of a new pnictide oxide having chromium-containing perovskite block evidenced the B site cation in the perovskite layer is not limited to Sc. This fact vastly expands possibility of new materials in pnictide oxide systems having perovskite-type oxide layer. We have already noticed structural similarity between copper sulfide oxides and iron pnictide oxides. In fact, as many as ten ions of B site cation in the perovskite-type oxide layer are reported in copper sulfide oxides. Therefore, there must be other elements, which can occupy the B site of perovskite-type layer in the pnictide oxides, if they satisfy following restrictions. At first, the cation must have appropriate valence and ionic radius. For example, valence of B site cation should be basically trivalent in the 22426 compounds and divalent in the hypothetical (Fe$_2$$Pn$$_2$)(Sr$_2$$M$O$_2$) compounds. Concerning ionic radii, ions with the radii from 61.5pm (Cr$^{3+}$, 6 coordination)[9] to 83.0pm (Mn$^{2+}$, 6 coordination, high spin)[8] are reported in cupper sulfide oxides. Existence of Cr-22426 and (Mn$_2$As$_2$)(Ba$_2$MnO$_2$)[7] indicates the restriction of ionic radii is almost same in copper sulfide oxides and pnictide oxides. Secondly, the appropriate valence of the B site cation and Fe$^{2+}$ at the pnictide layer should coexist under the synthesis conditions. Lastly, the B site cation should be introduced to the oxide layer prior to the pnictide layer to form iron pnictide layer. Narrow stable conditions of Fe$^{2+}$ compared to that of Cu$^{1+}$ and the last limitation might severely restrict the candidate elements. Further investigations of the novel system will be needed to clarify the actual criterions for these restrictions.

**Conclusions**

New layered iron pnictide oxides $M$-22426; (Fe$_2$As$_2$)(Sr$_4$$M$$_2$O$_6$) ($M$ = Sc, Cr) have been synthesized and their crystal structures were determined. Both materials consist of alternate stacking of anti-fluorite Fe$_2$As$_2$ and perovskite Sr$_4$$M$$_2$O$_6$ layer with the space group of *P4/nmm*. Interlayer Fe-Fe distance of 15.809 Å for Sc-22426 is the longest among the reported values for iron pnictides. In both magnetization and resistivity measurements, these new compounds did not exhibit superconducting transitions probably due to insufficient carrier concentration. Further attempts to introduce effective carriers to these materials must be promising for achieving high-$T_c$ superconductivity. On the other hand, existence of (Fe$_2$As$_2$)(Sr$_4$Cr$_2$O$_6$) suggested large variety of constituent elements in the perovskite-type oxide layer. Moreover, trends of local pnictide layer in pnictide oxides with perovskite-type oxide layer are different with other structural systems, such as 1111. Structural and chemical variations in perovskite layer will make this type of pnictide oxide as a repository of new functional materials including superconductors.

**Acknowledgement**

This work was performed under the inter-university cooperative research program of the Advanced Research Center of Metallic Glasses, Institute for Materials Research, Tohoku University. This work was partly supported by Center for Nano Lithography & Analysis, The University of Tokyo, supported by the Ministry of Education, Culture, Science and Technology (MEXT), Japan as well as a Grant-in-Aid for Young Scientists (B) no. 21750187, 2009.

**Figure captions**

Figure 1. Rietveld refinement of powder XRD and the crystal structure of Sc-22426(a): dashed line indicates calculated fit pattern and solid line indicate difference of observed and calculated pattern. Bars show diffraction peak positions of Sc-22426. Rietveld refinement of powder XRD of the Cr-22426(b): upper and lower bars show diffraction peak positions of FeAs and Cr-22426, respectively.

Figure 2. Bright-field TEM images and corresponding electron diffraction patterns of a Sc-22426 crystal viewed from the [1 -1 0] direction

Figure 3. Temperature dependence of ZFC and FC magnetization curves of the Sc-22426 and Cr-22426 bulk samples measured under 10 kOe.

Figure 4. Temperature dependences of the resistivity for the Sc-22426 and Cr-22426 bulk samples. Close-up of the resistivity curve for Cr-22426 is shown in the inset.

Figure 5. Relationship between $\alpha$ angle and *a* axis length (a) and Fe-*Pn* distance and *a* axis length (b) at room temperature except NdFeAsO(at 175 K). The values are obtained by undoped samples except TbFeAsO$_{0.9}$F$_{0.1}$. Closed symbols indicate iron pnictide oxides with perovskite-type oxide layer. Solid and dashed lines are guideline for eyes.

Table 1. Fitting parameters, bond lengths and bond angles for (Fe$_2$As$_2$)(Sr$_4$*M*$_2$O$_6$)

Figure 1

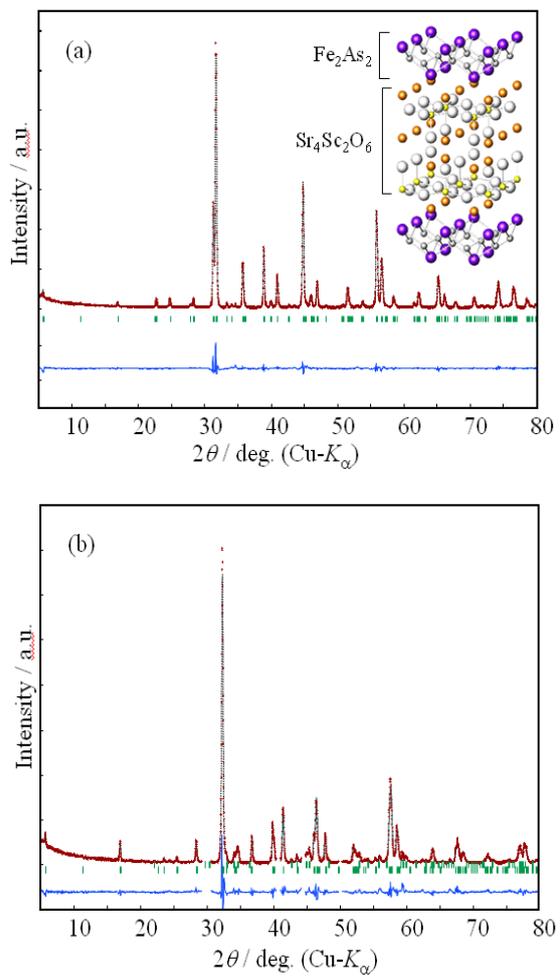

Figure 2

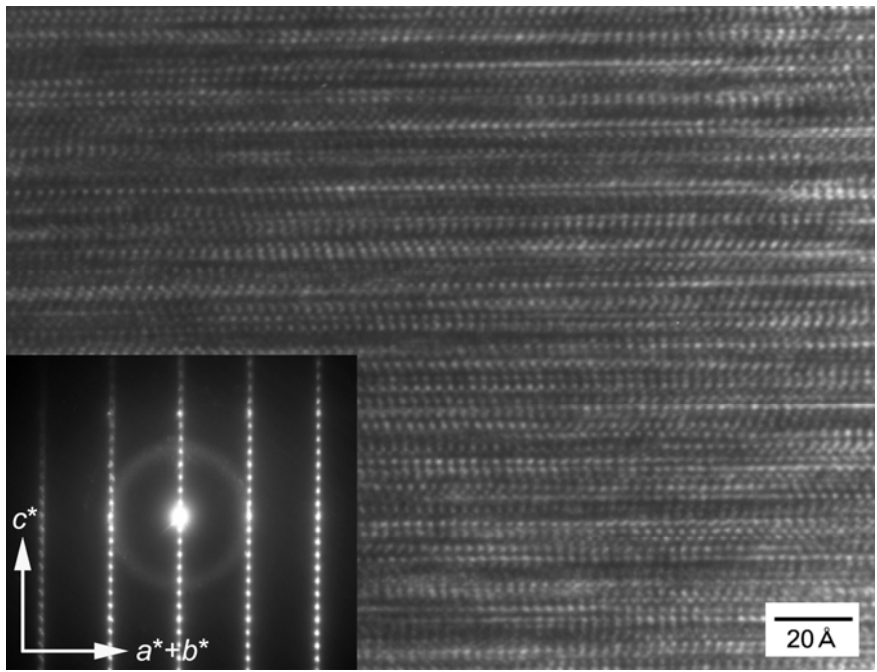

Figure 3

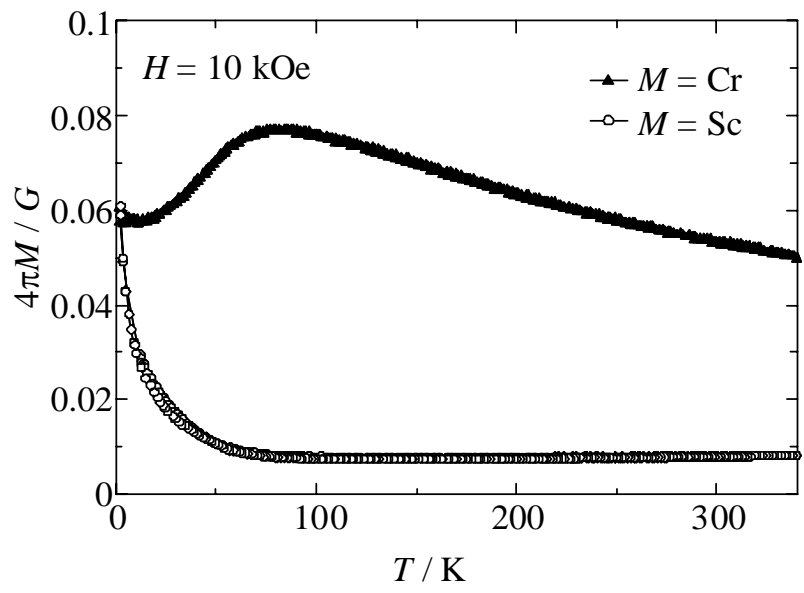

Figure 4

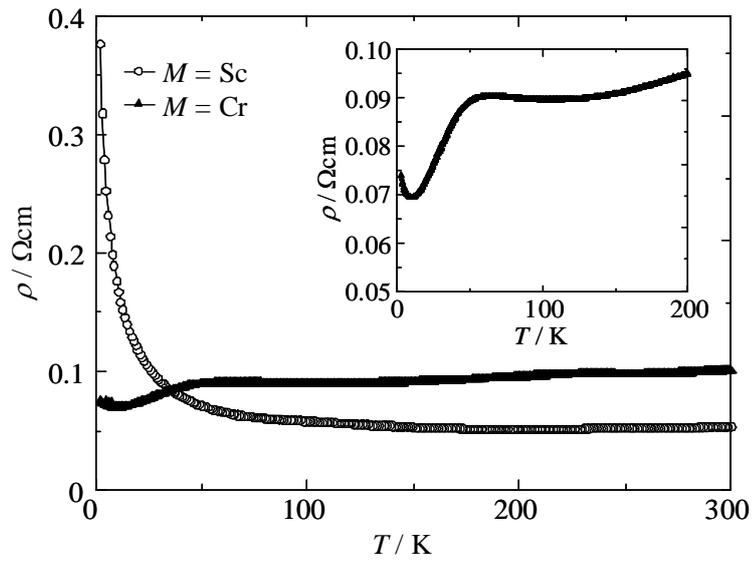

Fig. 5

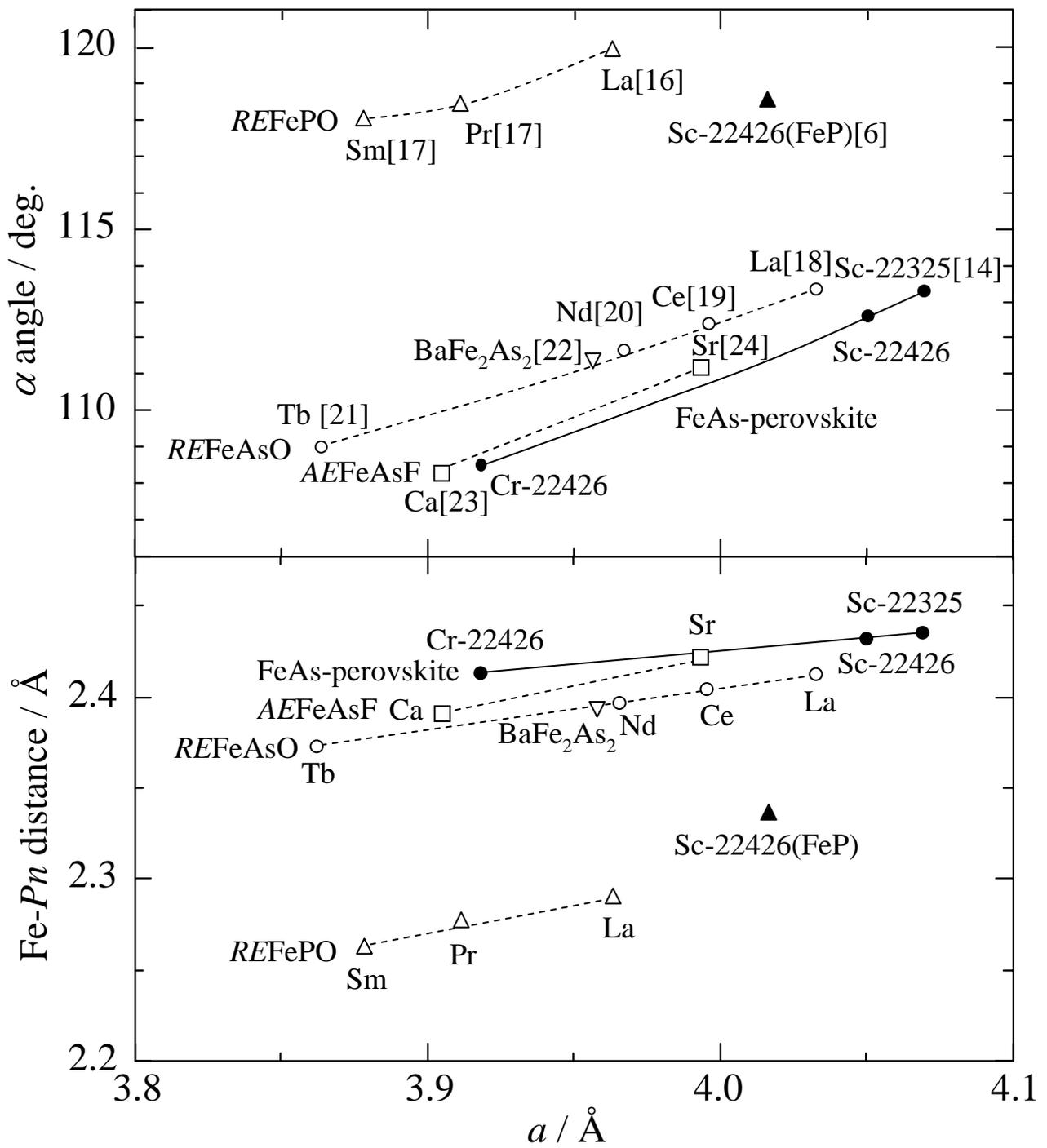

| Atom | x | y | z (Sc) | z (Cr) |
|---|---|---|---|---|
| Sc | 0.250 | 0.250 | 0.3071 | 0.3116 |
| Fe | 0.250 | -0.250 | 0.0000 | 0.0000 |
| Sr1 | -0.250 | -0.250 | 0.1887 | 0.1947 |
| Sr2 | -0.250 | -0.250 | 0.4153 | 0.4157 |
| O1 | 0.250 | -0.250 | 0.2857 | 0.2945 |
| O2 | 0.250 | 0.250 | 0.4301 | 0.4250 |
| As | 0.250 | 0.250 | 0.0854 | 0.0898 |

$R_{wp}$ = 9.32 $R_p$ = 6.48 $R_e$ = 9.70 $S$ = 0.5436 $R_F$ = 2.21 (Sc)
$R_{wp}$ = 13.68 $R_p$ = 9.99 $R_e$ = 9.98 $S$ = 1.3709 $R_F$ = 2.71 (Cr)

| | Sc | Cr |
|---|---|---|
| $\alpha$ | 112.6° | 108.5° |
| $\beta$ | 107.9° | 110.0° |
| $\gamma$ | 72.1° | 70.1° |
| Fe-As | 2.433 Å | 2.414 Å |
| Fe-Fe | 2.863 Å | 2.770 Å |

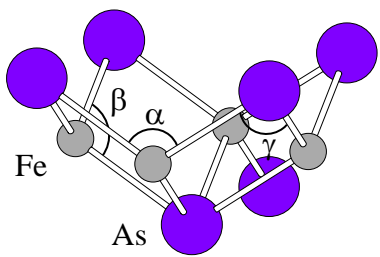